\documentclass[aps]{JHEP3}

\usepackage{amsfonts}

\newcommand{\na}{\nabla}
\newcommand{\ep}{\epsilon}

\newcommand{\vphi}{\varphi}
\newcommand{\lraw}{\longrightarrow}

\newcommand{\pa}{\partial}

\newcommand{\beq}[1]{\begin{eqnarray}\label{#1}}
\newcommand{\eeq}{\end{eqnarray}}
\newcommand{\nettwo}{{\cal N}=2}
\newcommand{\netone}{{\cal N}=1}

\title{Intersecting branes and adding flavors to the Maldacena-N\u{u}nez background}

\author{Xiao-Jun Wang\\
Interdisciplinary Center for Theoretical Study \\
University of Science and
Technology of China, \\
AnHui, HeFei, 230026, P.R. China\\
E-mail: wangxj@ustc.edu.cn}

\author{Sen Hu\\Interdisciplinary Center for Theoretical Study
and Department of Mathematics \\ University of Science and
Technology of China, \\
AnHui, HeFei, 230026, P.R. China\\
E-mail: shu@ustc.edu.cn}

\abstract{Following a proposal in ref.\cite{KK02}, we study adding
flavors into the Maldacena-N\u{u}nez background. It is achieved by
introducing spacetime filling D9-branes or intersecting D5$'$-branes into
the background with a wrapping D5-brane. Both D9-branes and D5$'$-branes
can be spacetime filling from the 5D bulk point of view. At the probe
limit it corresponds to introducing non-chiral fundamental flavors into
the dual $\netone$ SYM. We propose a method to twist the fundamental
flavor which has to involve open string charge. It reflects the fact that
coupling fundamental matter to SYM in the dual string theory corresponds
to adding an open string sector.}


\begin{document}

\section{Introduction}

It is expected that 4d pure Yang-Mills theory, with gluon as degrees of
freedom, is mapped to an entirely 5d non-critical closed string theory
background\cite{Polyakov99}. This expectation has been realized in many
supersymmetric examples starting from a proposal on AdS/CFT
correspondence\cite{Mald98}. In these examples we knew the geometry of the
background. Although it is difficult to quantize string theory on the
background, it was conjectured that at the large $N$ limit, the super
Yang-Mills theories (SYM) are dual of the low energy effective theories of
a closed string theory, i.e., supergravity(SUGRA), on various backgrounds.
When we add massless fundamental flavors to SYM, however, it is very
difficult to get the dual background. Adding fundamental flavors
effectively introduces an open string sector to the string theory. Light
open string states inevitably modify the original closed string
background. In other words, D-branes, which generate non-trivial
background, have to carry open string charge besides the closed string
charge (R-R charge).

It was noted by Karch and Katz\cite{KK02} that this difficulty can be
partially overcome at certain probe limit. They proposed an effective
method to add flavors into AdS/CFT correspondence via introducing few
D7-brane probe into $AdS_5\times S^5$. The key points of their proposal
can be summarized as follows:
\begin{itemize}
\item Adding fundamental flavors in the gauge theory is equivalent to
adding space-time filling D-branes into the 5d bulk theory.
\item It is not necessary to introduce orientifold in order to cancel
the R-R tadpole of space-time filling D-branes. Space-time filling
D-branes can wrap on a topological trivial cycle inside the
original background. This way they may not carry any charge and
avoid a R-R tadpole.
\item The stability of these space-time filling D-branes is ensured by
the negative mass modes in the AdS space, but the mass is above
the BF bound\cite{BF82a,BF82b} and they do not lead to an
instability in the curved 5d geometry.
\item At the probe limit, the effect of D-branes on the bulk
geometry can be ignored. The bulk geometry is still AdS but the
dual field theory is now CFT with a defect or boundary (dCFT).
When fundamental flavors get mass, it corresponds to bulk geometry
receiving supergravity fluctuations.
\end{itemize}
It is interesting to extend their proposal because it provides a possible
way to add fundamental flavors in gauge/string duality on known
backgrounds. Authors of ref.\cite{SS03} has incorporated massive
fundamental quarks in supergravity dual of $\netone$ SYM by introducing D7
brane probes in Klebanov-Strassler background. In ref.\cite{BEEGK03} it
was shown how to realize chiral symmetry breaking and production of
pseudoscalar mesons in some non-supersymmetric gauge/gravity duals. The
purpose of the present paper is to consider adding fundamental flavors
into $\netone$ SYM which is dual to the well-known Maldacena-N\u{u}nez
(MN) background\cite{MN01}.

The MN background is described by N D5-branes wrapping on a
topological non-trivial 2-cycle inside a CY three-fold. The normal
bundle within the CY space has to be twisted in order to preserve
some supersymmetry. This configuration is much complicated than
the one considered by Karch and Katz, in which the D3-branes
background are flat. Similar to adding D7-branes into the D3-brane
background in the AdS case, we may consider to add finite number
of D9-branes or D5-branes into the MN background at the large N
limit respectively. In the following we shall discuss two
different cases respectively.
\begin{itemize}
\item[{\bf A}] Adding M D9-branes: In the flat case, D5-D9 system supports
$D=6,\,\netone$ supersymmetry. Adding D9-branes is effectively to
add M flavor hypermultiplets into the 6d $SU(N)$ gauge theory.
When D5-branes wrap on a supersymmetric 2-cycle, the 6d $\nettwo$
vector multiplet on the world-volume of D5-branes reduces to 4d
$\nettwo$ or $\netone$ vector multiplet. The background geometry
is $\mathbb{R}^4\times CY_3$. Therefore, D9-branes have to wrap on
a CY three-fold so that supersymmetry is still kept. This
configuration, however, seems to receive some unexpected features.
The first problem should be asked whether D9-branes can be treated
as a probe since CY three-fold is open. If it is not true, MN
solution is not a well-defined background to describe this system.
The second problem is that the D9-branes shall always carry
induced R-R charge from the background, whether they are treated
as probe or not. This is different from the case discussed in
\cite{KK02}. At rigid probe limit the effects of R-R tadpole can
be ignored. The original MN background is still good description
on this configuration. When we study this system beyond probe
limit, we encounter dangerous R-R tadpoles. They have to be
cancelled by either introducing orientifold planes or anti-D9s.
The latter is in general unstable, and the former destroy MN
background description.

Although there are those unpleasant problems, it is helpful to
count spectrum of the dual field theory. The twist on vector
multiplet by D5-branes is well-known. The twist on hypermultiplet,
however, should be checked carefully. In general, the original
twist on 6d hypermultiplet is no longer to leave any massless
states, otherwise D9-branes can not be felt by the field theory.
As discussed above, when massless fundamental flavors are
introduced, D5-branes will carry open string charge. A natural
suggestion is that these charges should play a role in twisting
supermultiplet. We will show that it indeed works.
\item[{\bf B}] Adding M D5$'$-branes: The D5$'$-branes intersect with
the original 5-branes, and have four dimensions in common. In the
flat case, this system supports a $D=4,\,\nettwo$ $SU(N)\times
SU(M)$ gauge theory. Adding D5$'$-branes is to add M flavor
hypermultiplet into the 4d $SU(N)$ gauge theory. When D5-branes
wrap on a supersymmetric 2-cycle inside a CY 3-fold, the
D5$'$-branes should wrap another orthogonal supersymmetric
2$'$-cycle inside the CY space. If the 2$'$-cycle is topological
non-trivial, D5$'$-branes carry R-R charge from the background.
The charge, however, does not bring dangerous tadpole problem
since the CY space is open while the 2$'$-cycle is compact. If the
2$'$-cycle is topological trivial, D5$'$-branes do not get R-R
charge and we don't have tadpole problem again. Thus adding
D5$'$-branes is a nice choice: D5$'$-branes can be treated as a
probe in the original D5-brane background and it does not cause
the tadpole problem.

The twist on various supermultiplet, similar to the D5-D9 case,
should be checked carefully. The open string charge will play a
role in twisting in the hypermultiplet.
\end{itemize}

When D5$'$-branes does not carry induced charge from the
background, stability of D-branes should be considered. The
stability of D-branes is ensured by the similar mechanism
in\cite{KK02,KR01,BP03}: there are negative mass modes in 5d world
which control the slipping of the D-branes off the cycle they
wrap, but the mass is not negative enough to lead to an
instability in the curved 5d geometry. We will show that similar
mechanism works in our cases. The negative mass modes in 5d bulk
spacetime are generated by Kaluza-Klein spectrum on a compact
5-cycle inside a CY three-fold, and have a lower bound (similar to
BF bound in AdS space).

The paper is organized as follows: In section 2 we consider adding
D9-branes to the MN background. By a careful twist, the flat part of the
system supports $\netone$ SYM with fundamental matter. In section 3 we
show how to add D5$'$-branes to the MN background and twist it. The flat
part of this system again supports $\netone$ SYM with fundamental matter.
We give a brief conclusion in section 4. The appendix of this paper
devotes to computing KK modes on the compact part of CY three-fold which
is described by the MN solution.

\section{Adding D9-branes into the MN background}

\subsection{The flat spectrum}

The massless spectrum for flat D5-D9 system has been given in
ref.\cite{PbookII}. In the following we review some details.
Assuming D5-branes spread in \{0, 1, 2, 3, 4, 5\} directions, the
massless spectrum of this system is as follows,
\begin{itemize}
\item 5-5 open strings generate $D=6,\,(1,1)$ vector supermultiplet with
R-symmetry $SO(4)\simeq SU(2)_L\times SU(2)_R$.
\item 9-9 open strings generate $D=10,\,\netone$ vector supermultiplet.
\item 5-9 open strings: In the NS-sector four periodic world-sheet fermions
$\psi^i$, namely in ND directions $i=6,7,8,9$, generate four
massless bosonic states. We label their spins in the (6,7) and
(8,9) planes, $|s_3,s_4\rangle$ with $s_3,\,s_4$ taking values
$\pm\frac{1}{2}$. The GSO projection requires $s_3=s_4$ so that
two massless bosons survive. In the R-sector four transverse
world-sheet fermions $\psi^m$ with $m=2,3,4,5$ are periodic. They
again generate four massless fermionic states, and are labelled by
$|s_1,s_2\rangle$. The GSO projection requires
$s_1=-s_2=\pm\frac{1}{2}$ so that again two massless fermions
survive. Then massless content of the 5-9 spectrum amounts to a
$D=6$ half-hypermultiplet. The other half comes from strings of
opposite orientation, 9-5 strings. From the point of view of
D9-brane world volume, 5-9 and 9-5 carry opposite charges. In
other words, they associate to a global $U(1)_o$ symmetry, where
$``o"$ denotes orientation \footnote{This global $U(1)_o$
symmetry, however, is $U(1)$ gauge symmetry on D9 world-volume
from view point of world-volume field on D9-branes.}. For the sake
of convenience we split 10d Lorentz group into \beq{2.1}
SO(1,9)&\lraw& SO(1,1)\times SO(4)_N\times SO(4)_D \nonumber \\
  &\simeq &SO(1,1)\times SU(2)_{\rm NL}\times SU(2)_{\rm NR}\times
  SU(2)_{\rm DL}\times SU(2)_{\rm DR}.
\eeq
The massless states lie in the following representation of the group
$$SO(1,1)\times SU(2)_{\rm NL}\times SU(2)_{\rm NR}\times
  SU(2)_{\rm DL}\times SU(2)_{\rm DR}\times U(1)_o,$$

bosons: (0, 1, 1, 2, 1, $\pm$)

fermions: ($\pm\frac{1}{2}$, 1, $2'$, 1, 1, $\pm$)
\end{itemize}

\subsection{Wrapping D5-branes on a 2-cycle}

Wrapping world-volume of N D5-branes on a topological non-trivial
2-cycle leads to a non-supersymmetric 3-branes. In order to
preserve supersymmetry one need to twist the normal bundle of this
2-cycle, i.e., to identify $U(1)$ charge on a 2-cycle (denotes by
$U(1)_J$) with an external gauge field. For D5-branes there are
two choices: to identify the $U(1)_J$ charge with the charge of
diagonal sub-$U(1)_D$ group of R-symmetry group $SU(2)_{\rm
DL}\times SU(2)_{\rm DR}$, or with the charge of subgroup
$U(1)_{\rm DL}(U(1)_{\rm DR})$ of $SU(2)_{\rm DL}(SU(2)_{\rm
DR})$. Some massless states become massive under the twist and
decouple with other massless states. The flat part of D5-branes
supports a (3+1)-dimensional twist gauge theory with $\nettwo$ or
$\netone$ supersymmetry corresponding to the previous two choices.

When M D9-branes are introduced, the twist on vector
supermultiplet on D5-branes is the same as the case without
D9-branes. The key point is to focus on twisting hypermultiplet
generated by 5-9 and 9-5 string. When an D5-brane wrap on $S^2$,
the 6d hypermultiplet in the previous subsection reduces to:
\begin{itemize}
\item[] Bosons: $(2,1,\pm)\times|0,0\rangle_0$.
\item[] Fermions: $(1,1,\pm)\times \left(|\pm\frac{1}{2},\frac{1}{2}\rangle_-\oplus
   |\pm\frac{1}{2},-\frac{1}{2}\rangle_+\right)$ ,
\end{itemize}
where $|s_0,s_1\rangle$ is the spin basis in 4d spacetime,
subscript $``0,\,\pm"$ denotes the $U(1)_J$ charge, and $({\rm x,
x, x})$ denotes the representation of $SU(2)_{\rm DL}\times
SU(2)_{\rm DR}\times U(1)_o$.

Now let us consider twisting on the above states. Naively we may
consider two choices mentioned above:
\begin{itemize}
\item[1)] $U(1)_J=U(1)_D\in D(SU(2)_{\rm DL}\times SU(2)_{\rm DR})$.
\begin{itemize}
\item[] Bosons: $2\times |0,0\rangle_{\pm}$.
\item[] Fermions: $2\times |\pm\frac{1}{2},\frac{1}{2}\rangle_-\oplus
       2\times |\pm\frac{1}{2},-\frac{1}{2}\rangle_+$.
\end{itemize}
Here subscript $``\pm''$ denotes the total $U(1)=U(1)_J\times
U(1)_D$ charge. Hence there are no massless states surviving and
the resulted field theory is still pure $D=4,\,\nettwo$ SYM.
\item[2)] $U(1)_J=U(1)_L\in SU(2)_{\rm DL}$. This choice is the same as the
first one, but the resulted field theory is pure $D=4,\,\netone$
SYM.
\item[3)] $U(1)_J=U(1)_R\in SU(2)_{\rm DR}$. Denoting the total charge by
$U(1)=U(1)_J\times U(1)_R$ we have
\begin{itemize}
\item[] Bosons: $4\times |0,0\rangle_{0}$.
\item[] Fermions: $2\times |\pm\frac{1}{2},\frac{1}{2}\rangle_-\oplus
       2\times |\pm\frac{1}{2},-\frac{1}{2}\rangle_+$.
\end{itemize}
Then this choice breaks supersymmetry.
\end{itemize}

We see that all of the above choices do not generate massless
fundamental matters in the framework of 4d supersymmetric gauge
theory. It can be understood naturally: The fundamental matter
carries open string charges. The charges should play a role in
twisting on fundamental matter. Consequently we can say that the
4d field theory gets effects from the open string sector.
Therefore, we propose the following twist,
\begin{itemize}
\item[1)] $U(1)_J=U(1)_D\in D(SU(2)_{\rm DL}\times SU(2)_{\rm DR})=
   U(1)_o$. The total charge is of $U(1)=U(1)_J\times U(1)_D\times
   U(1)_o$.
\begin{itemize}
\item[] Bosons: $2\times |0,0\rangle_0\oplus |0,0\rangle_{++}\oplus
|0,0\rangle_{--}$.
\item[] Fermions: $ |\pm\frac{1}{2},\pm\frac{1}{2}\rangle_0\oplus
   |\pm\frac{1}{2},\frac{1}{2}\rangle_{--}\oplus |\pm\frac{1}{2},
   -\frac{1}{2}\rangle_{++}$.
\end{itemize}
Two real scalars (one complex scalar) and one Majorana spinor
survive under the twist. They form a chiral multiplet of
$D=4,\,\netone$ superalgebra. Because the supermultiplet generated
by 5-5 string does not carry $U(1)_o$ charge, the twist on vector
multiplet on D5-branes is not changed. Then we obtain
$D=4,\,\netone$ gauge theory which contains a vector multiplet, a
chiral multiplet in the adjoint representation and M chiral
multiplets in the fundamental representation of the gauge group.
\item[2)] $U(1)_J=U(1)_L\in SU(2)_{\rm DL}=U(1)_o$. Again two
real scalars and one Majorana spinor survive under the twist. We
obtain a $D=4,\,\netone$ gauge theory which contains a vector
multiplet and M chiral multiplets in the fundamental
representation of the gauge group.
\item[3)] $U(1)_J=U(1)_R\in SU(2)_{\rm DR}=U(1)_o$. No massless
states survive at this case.
\end{itemize}

Our proposal is indeed valid for 1) and 2). The $U(1)_o$ charges
play essential role in twisting on various string spectrum.
Because the $U(1)_o$ symmetry is gauge symmetry on D9-brane
world-volume, it looks like that some world-volume field are
turned on. This is true since endpoints of each 5-9 and 9-5 string
pair carry opposite charges. Then a non-vanishing flux crosses on
a pair of endpoints and lies within D9-brane worldvolume. The
fluxes act as background gauge field strength $F$ on world-volume.
Similar to role of $B$-field background discussed in
\cite{Witten02}, the worldvolume $F$-field compensates breaking of
supersymmetry when introducing D9-brane in wrapping D5-brane
background. The same conclusion can be obtained from the following
supergravity argument.

Because endpoints of an open string couple to NS 1-form potential,
NS 1-form potential should be introduced into supergravity when we
introduce fundamental flavor in the dual field theory. To make
$S^3$ reduction, R-R 2-form potential induces $SO(4)$ gauge
field\cite{CLP00} in the 7d gauged supergravity while NS 1-form is
still kept. Then the twist on normal bundle of the 2-cycle now
becomes \beq{2.2}\omega_M=A_M^{RR}+A_M^{NS}, \eeq where $\omega_M$
is the spin connection of the 2-cycle. The connection on normal
bundle is now cancelled by R-R field generated from the close
string sector together with NS 1-form generated by the open string
sector.

\subsection{Discussions}

Several open questions should be asked here. The first question is
whether an D9-brane can be treated as a probe in the wrapped
D5-brane background for $N\rightarrow\infty,\,M\sim$ fixed. If we
treat an D9-brane as a probe, i.e., the effect of D9-branes is
ignored, D9-branes wrap on the CY three-fold described by the
wrapping D5-brane background. The energy of D9-branes, however, is
still infinity since CY space is open. The same puzzle appears in
introducing a D7-branes probe into a D3-brane background proposed
by Karch and Katz\cite{KK02}. An effective treatment is to
introduce a cut-off near the boundary. Then energy of D7-branes
will be of order to the energy of a D3-brane probe locating at the
boundary of the background geometry. Consequently D7-branes can be
treated as a probe as same as a D3-branes probe. For D9-branes,
however, this treatment does not work well. For example, in the MN
background the ratio of the energy of D9-branes to one of the
D5-brane background locating at the boundary is of order
$M\ln^4{(r_0/l)}/N$ where $r_0\rightarrow\infty$ is a cut-off and
$l$ is the definite scale. Therefore, it is decided by fine tuning
between $M/N$ and $\ln^4{(r_0/l)}$ whether D9-branes can be
treated as a probe.

Whether D9-branes are treated as probe or not, D9-branes carry R-R
charge induced from wrapped D5-brane background. At rigid probe
limit the analysis is fine. Beyond probe limit, however, we
encounter unexpect R-R tadpole problem. This correction is order
$M/N$, and is also implied by fundamental matter spectrum found in
previous subsection that the spectrum is anomalous due to lack of
anti-fundamental matter. The anomaly (or R-R tadpole) makes
theories be inconsistent. They should be cancelled via introducing
orientifold planes or anti-D9s. The former will destroy original
wrapped D5-brane background even though D9-branes are treated as
probe. The introduction on anti-D9-branes looks like strange since
brane-anti-brane system is in general non-supersymmetric and
unstable. This is not fact when background world-volume $F$-field
is turned on. It was shown in refs.\cite{BK02,BOS02} that, the
brane-anti-brane system will be 1/4 BPS states when background
magnetic fields take opposite directions on worldvolume of branes
and anti-branes respectively. In addition, the specific
worldvolume background electric field makes the usual tachyonic
degrees disappear. Thus the system will be stable. Both of these
conditions can be satisfied here since a non-vanishing background
$F$-field is turned on D9-brane worldvolume. Then introduction on
anti-D9s to cancel R-R tadpole is good choice. The open strings
stretching between D5-branes and anti-D9s give massless
anti-fundamental flavors which cancels anomaly in massless
spectrum.

When D9-branes are treated as a probe, it is particularly
interesting for the second choice on twist. For this twist
supergravity on the MN background is dual to a (non-chrial)
$SU(N)$ SQCD with M fundamental flavors at the probe limit and the
large $N$ limit. It reflects the fact that at the large $N$ limit
the role of gluons is more important than the role of quarks
(without chiral symmetry spontaneously breaking effects). The
effect of fundamental flavors is reflected by DBI action on
D9-brane worldvolume as in ref.\cite{meson1,meson2}. In order to
manifest the quantum number of fundamental flavors, NS gauge field
should be turned on in supergravity, i.e., we have to consider
type I string fluctuations on the type IIB background. It
corresponds to fluctuation on the MN background. The leading order
correction is expected to be order of $M/N$ but cancelled by
anti-D9s.

\section{Adding intersecting D5$'$-branes}

We consider $N$ ($N\to\infty$) D5-branes intersecting with $M$
($M\sim$ fixed) D5$'$-branes. The D5-branes spread in
\{0,1,2,3,4,5\} directions while the D5$'$-branes spread in
\{0,1,2,3,6,7\} directions. It is different from the D5-D9 system.
We can naturally treat D5-branes as a background and D5$'$-brane
as a probe on this background.

\subsection{Flat spectrum}

The flat spectrum on intersecting D5-branes was presented in
ref.\cite{BDL96}. Here we list the main results for the sake of
convenience as follows. Massless spectrum of 5-5 string and
5$'$-5$'$ string are 6d (1,1) vector supermultiplet on D5-brane
and D5$'$-brane worldvolume respectively. The new ingredient is
the 5-5$'$ string. We divide 10d spacetime into three sectors: the
NN sector with Newmann-Newmann boundary conditions for $X^\mu$
($\mu=0,1,2,3$), the DD sector with Dirichlet-Dirichlet boundary
condition for $X^m$ ($m=8,9$) and the ND sector with
Dirichlet-Newmann boundary condition for $X^i$ ($i=4,5,6,7$). In
the NS sector, two real scalars survive by GSO projection, namely
$|s_2,s_3\rangle$ with $s_2=s_3=\pm 1/2$ where $s_2$ and $s_3$
denotes spins in (4,5) and (6,7) planes. In the R sector, two
fermions survive by GSO projection. That is $|s_1,s_4\rangle$ with
$s_1=-s_4=\pm 1/2$, where $s_1$ and $s_4$ denotes spins in (2,3)
and (8,9) planes respectively. Together with massless states
generated by 5$'$-5 string, they form a hypermultiplet of
$D=4,\,\nettwo$ superalgebra.

Ten-dimensional Lorentz symmetry now breaks down to
$$SO(1,9)\lraw SO(1,3)\times SO(2)_{\rm ND}\times SO(2)_{\rm ND'}\times
SO(2)_{\rm D},$$ where ``ND" denotes ND directions on D5-brane
worldvolume and ``ND$'$" denotes ND directions on D5$'$-brane
worldvolume. Together with 4d R-symmetry, the total symmetry now
is
$$SO(1,3)\times U(1)_{\rm ND}\times U(1)_{\rm ND'}\times
U(1)_{\rm D}\times U(1)_o$$.

The representations of various massless field are as follows,
\begin{itemize}
\item Six-dimensional vector multiplet generated by 5-5 string:
\begin{itemize}
\item[] $V_M$: $(4,0,0,0,0)\oplus (1,\pm,0,0,0),$
\item[] $\phi^A$: $(1,0,\pm,0,0)\oplus (1,0,0,\pm,0),$
\item[] $\psi^+$: $(2,+,+,+,0)\oplus (2,+,-,-,0)\oplus (\bar{2},-,+,+,0)
   \oplus (\bar{2},-,-,-,0),$
\item[] $\psi^-$: $(2,-,+,-,0)\oplus (2,-,-,+,0)\oplus (\bar{2},+,+,-,0)
   \oplus (\bar{2},+,-,+,0).$
\end{itemize}
\item Four-dimensional hypermultiplet generated by 5-5$'$ string:
\begin{itemize}
\item[] Scalars $\vphi$ : $(1,+,+,0,\pm)\oplus (1,-,-,0,\pm),$
\item[] Spinors $\lambda$: $(0,0,+,\pm)\times |s_0,\frac{1}{2}\rangle\,\oplus\,
                  (0,0,-,\pm)\times |s_0,-\frac{1}{2}\rangle,$
\end{itemize}
where $|s_0,s_1\rangle$ denotes spinor in (3+1)-dimension spacetime.
\end{itemize}

\subsection{Twisting on spectrum}

Now we consider D5-branes wrapping on a supersymmetric 2-cycle. It
twist the above spectrum and consequently some states become
massive and decouple from others. As for D5-D9 system, open string
charge $U(1)_o$ plays an important role in twisting the
fundamental hypermultiplet. We claim that the following two
choices are valid for our purpose.
\begin{itemize}
\item[1)] $U(1)_{\rm ND}=U(1)_{\rm D}=U(1)_o$. Total charge is defined by
$U(1)=U(1)_{\rm ND}\times U(1)_{\rm D}\times U(1)_o$. Then under
$SO(1,3)\times U(1)$ symmetry various states reduce to
\begin{itemize}
\item[] $V_M$: $4_0\oplus 1_\pm,$
\item[] $\phi^A$: $2\times 1_0\oplus 1_\pm,$
\item[] $\psi^+$: $2_0\oplus \bar{2}_0\oplus 2_{++}\oplus \bar{2}_{--},$
\item[] $\psi^-$: $2_0\oplus \bar{2}_0\oplus 2_{--}\oplus \bar{2}_{++},$
\item[] $\vphi$: $\,\,2\times 1_0\oplus 1_\pm,$
\item[] $\lambda$: $\,\,|s_0,s_1\rangle_0\,\oplus\,
    |s_0,\frac{1}{2}\rangle_{++}\,\oplus\,|s_0,-\frac{1}{2}\rangle_{--},$
\end{itemize}
Therefore, we end with the $D=4,\,\netone$ $SU(N)$ gauge theory
whose massless spectrum includes a vector multiplet, a chiral
multiplet in the adjoint representation and M chiral multiplets in
the fundamental representation of the gauge group. When
D5$'$-branes are absent, this choice reduces to $U(1)_J=U(1)_D\in
D(SU(2)_L\times SU(2)_R)$, and the resulted field theory is
$D=4,\,\nettwo$ pure SYM.
\item[2)] $U(1)^{1/2}_{\rm ND}=U(1)_{\rm ND'}=U(1)_{\rm D}=U(1)_o$. Total
charge is defined by $U(1)=U(1)_{\rm ND}\times U(1)_{\rm ND'}\times
U(1)_{\rm D}\times U(1)_o^{3/2}$. Under $SO(1,3)\times U(1)$ symmetry
various states reduce to
\begin{itemize}
\item[] $V_M$: $4_0\oplus 1_\pm,$
\item[] $\phi^A$: $2\times 1_+\oplus 2\times 1_-,$
\item[] $\psi^+$: $2_0\oplus \bar{2}_0\oplus 2_{++}\oplus \bar{2}_{--},$
\item[] $\psi^-$: $2\times 2_-\oplus 2\times\bar{2}_+,$
\item[] $\vphi$: $\,\,2\times 1_0\oplus 1_\pm,$
\item[] $\lambda$: $\,\,|s_0,s_1\rangle_0\,\oplus\,
    |s_0,\frac{1}{2}\rangle_{+}\,\oplus\,|s_0,-\frac{1}{2}\rangle_{-},$
\end{itemize}
We again end with the $D=4,\,\netone$ $SU(N)$ gauge theory but
whose massless spectrum now includes a vector multiplet and M
chiral multiplets in the fundamental representation of the gauge
group. When D5$'$-branes are absent, this choice reduces to
$U(1)_J=U(1)_L\in SU(2)_L$, and the resulted field theory is
$D=4,\,\netone$ pure SYM.
\end{itemize}

Most remarks on D5-D9 system in the previous section is still
applicable here. The main advantage for D5-D5$'$ system is
D5$'$-branes can be treated as a probe in the D5 background
without any fine tuning like in a D5-D9 system. Hence at the probe
limit the supergravity background on wrapping D5-branes can be
used to study the dual gauge theory with fundamental flavors. In
particular, supergravity on the MN background is dual to the large
N SQCD with (non-chiral) fundamental quarks. The effect of
fundamental flavors is reflected by the DBI action on a
D5$'$-brane probe. In order to manifest the quantum number of
fundamental flavors, NS gauge field should be turned on in
supergravity. It corresponds to fluctuation on the MN background
and is expected to be of order $M/N$.

Different aspect arises from considering a 2-cycle (denotes
2$'$-cycle) wrapped by D5$'$-branes. It is orthogonal to the
2-cycle wrapped by D5-branes and inside a CY three-fold. The
2$'$-cycle can be either topological trivial or non-trivial. If it
is topological non-trivial, D5$'$-branes will carry induced charge
from the D5-brane background. Because D5$'$-branes are not
spacetime filling now, there is no dangerous tadpole problem.
Meanwhile, D5$'$-branes are ``real" probes on background. They are
probe interaction not only of fundamental flavors but also of
gauge bosons on D5-branes. If the 2$'$-cycle is topological
trivial, D5$'$-branes do not carry any induced charge. Then there
is no R-R tadpole excitation even for spacetime filling
D5$'$-branes. The stability of D5$'$-branes is ensured by negative
mass modes propogating in 5d spacetime. In addition, absence of
R-R tadpole implies that there should be no anomaly in massless
spectrum. In other words, anti-fundamental flavors should appear
besides of fundamental flavors found previously. This is achieved
because although total induced R-R charge carried by D5$'$-branes
vanishes, the R-R potential does not vanish in general. When
D5$'$-branes wrap on topological trivial 2$'$-cycle inside CY
3-fold, the remainder part of CY space can be represented as
product form of two topological equivalent $S^2$. The R-R field
strength carried by D5$'$-branes can spread along two different
transverse directions in order to get vanishing total charge. A
special case is that one component spread on a $S^2$ wrapped by
D5-branes, and another component spread on another $S^2$. The flux
consequently splits $S^2$ into two disconnect pieces since $S^2$
is compact, i.e., D5-branes are also split into two pieces. The
two pieces carry opposite charge since they possess opposite
orientation. The open strings stretching between D5$'$-branes and
one piece of D5-branes give massless fundamental flavors, while
one stretching between D5$'$-branes and another piece of D5-branes
give massless anti-fundamental flavors. In other words, the
flavors are doubled when D5$'$-branes wrap on topological trivial
2$'$-cycle\footnote{The splitting, however, does not affect gauged
vector supermultiplet because that multiplet is supported by flat
part of D5-branes, contrary to hypermultiplet which associated to
wrapped directions of 5-branes.}. This point will be manifested
via geometry analysis in the next subsection.

\subsection{Geometry analysis}

We are interested in the MN background. At the probe limit the
effect of D5$'$-branes is ignored. The background reads\cite{MN01}
\beq{3.1}
   ds^2&=&e^\Phi dx^2_{1,3}+e^\Phi\alpha'g_sN\left[
   e^{2h}(d\theta_1^2+\sin^2{\theta_1}d\phi_1^2)
     +d\rho^2+\sum_{a=1}^{3}(w^a-A^a)^2\right], \nonumber \\
   F_{(3)}&=&2\alpha'g_sN\prod_{a=1}^3(\omega^a-A^a)
        +\alpha'g_sN\sum_{a=1}^3F^a\wedge\omega^a
\eeq
with $F^a=dA^a+\ep^{abc}A^b\wedge A^c$,
\beq{3.2}
   e^{2h}&=&\rho\coth{2\rho}-\frac{\rho^2}{\sinh^2{2\rho}}
             -\frac{1}{4},  \\
   e^{2\Phi}&=&e^{2k-2h}=ce^{-h}\sinh{2\rho}, \hspace{0.8in}
   a=\frac{2\rho}{\sinh{2\rho}}, \nonumber
\eeq
where $c$ is an integral constant, $w^a$ parameterize the 3-sphere,
\beq{3.3}
\omega^1&=&\frac{1}{2}(\cos{\psi}d\theta_2+\sin{\psi}\sin{\theta_2}d\phi_2)
  \nonumber \\
\omega^2&=&-\frac{1}{2}(\sin{\psi}d\theta_2
  -\cos{\psi}\sin{\theta_2}d\phi_2) \\
\omega^3&=&=\frac{1}{2}(d\psi+\cos{\theta_2}d\phi_2) \nonumber
\eeq and gauge field $A_a$ are written as\beq{3.4}
   A^1=-\frac{1}{2}ad\theta_1,\quad
   A^2=\frac{1}{2}a\sin{\theta_1}d\phi_1,\quad
   A^3=-\frac{1}{2}\cos{\theta_1}d\phi_1.
\eeq Then the D5-brane charge is given by
\beq{3.5}\tau_5=\int_{S^3}F_{(3)}\quad\sim\quad\alpha' g_sN. \eeq
D5-branes wrapping on $S^2$ is parameterized by
$\{\theta_1,\,\phi_1\}$ which shrinks to zero for $\rho\to 0$.
Then D5-branes disappear in resolution and fractional D3-branes
are created. D5$'$-branes, meanwhile, may wrap on another
orthogonal $S'^2$ parameterized by $\{\theta_2,\,\phi_2\}$ which
carries induce charge. This $S'^2$ is with finite radius for
$\rho\to 0$. Consequently no fractional D3$'$-branes are created
in resolution. The finite size of $S'^2$ implies that gauge theory
on probing D5$'$-branes flow to a conformal point in IR. Hence
effects of fundamental flavors are manifested in IR even at the
probe limit. If we go beyond the probe limit, both $S^2$ wrapped
by D5 and D5$'$-branes no longer shrink to zero because there is a
new $U(1)$ flux go through them. This $U(1)$ flux, however, does
not yield any singularity. It just reflects the effect of
fundamental matter in background and shifts charge carried by the
D5$'$-brane by M.

Another interesting possibility is that D5$'$-branes wrap on a
topological trivial 2-cycle inside a CY three-fold, namely
$\{\rho,\,\psi\}$ directions. Thus they are now spacetime filling
branes. To introduce a cut-off in UV (large $\rho$), the energy
ratio of D5$'$-branes to a D5-brane probe locating at UV is about
$(\ln\ln{r_0/l})^{-2}$. It means that D5$'$-brane can be
consistently treated as a probe. The induced magnetic and electric
charge carried by D5$'$-branes are now \beq{3.6}\int
F'_{(3)}&\sim& \alpha'g_s N(\cos{\theta_1}-\cos\theta_2),
 \nonumber \\
 \int *F'_{(3)}&\sim &\int(\tan\theta_2d\theta_2-\tan\theta_1d\theta_1)
  =0.
\eeq Charge quantization condition requires
\beq{3.7}\int
F'_{(3)}&=&0\quad\Rightarrow\quad \theta_1=\theta_2. \eeq
In other words,
D5$'$-branes do not carry any induced charge. The stability of
D5$'$-branes is ensured by the mechanism mentioned above.

The eq.~(\ref{3.6}) implies that induced magnetic and electric
flux carried by D5$'$-branes do not vanish although total charges
vanish. In particular, the electric field strength is proportional
to
 \beq{3.a} *F'_{(3)}&\sim & dx^0\wedge dx^1\wedge dx^2\wedge dx^3
   \wedge d\rho\wedge d\psi\wedge(\tan{\theta_1}d\theta_1
    -\tan{\theta_2}d\theta_2). \eeq
It indicates one components of electric flux spread in $\theta_1$
direction which splits $S^2$ wrapped by D5-branes into two pieces.
It is consistent with argument in previous subsection.
Consequently, we obtain not only M fundamental flavors, but also M
anti-fundamental flavors.

There is an additional remark on spacetime filling D5$'$-branes
because it is similar to the case described by Karch and
Katz\cite{KK02}. The simple induced form of the worldvolume metric
on D5$'$-branes is modified when the mass term for chiral
multiplet is turned on. It is achieved by separating D5-branes and
D5$'$-branes with a small distance c in $x^9$ direction. Using
parameterization of eq.~(\ref{3.3}), $x^9=e^{\Phi/2}\sin\theta_2$
in unit $\sqrt{\alpha'g_s N}$ leads to
\beq{3.8}\theta_2=\sin^{-1}(ce^{-\Phi(\rho)}/2)\sim
ce^{-\Phi(\rho)/2} +\cdots, \eeq where on the D5$'$ worldvolume
$\rho$ takes values between $\infty$ and $0$. For $c=0$ we recover
the topological trivial 2$'$-cycle defined by
$\theta_1=\theta_2=0$. Then fluctuation of worldvolume scalar
$\theta_2$ corresponds to the mass perturbation of the chiral
multiplet. The ingredient different from ref.\cite{KK02} is that
D5$'$ is till space-time filling instead of ending in the middle
of nowhere.

\section{Summary and more discussions}

We have shown that a few fundamental flavors can be introduced
into the closed string dual of large $N$ non-chiral SQCD by adding
a few D9-branes or orthogonal D5$'$-branes to a wrapped D5-brane
background. The twist on the resulted spectrum is much subtler
than without probing D-branes. The essential point is that an open
string sector is introduced in the string dual of the field
theory. Consequently open string charge plays a role in twisting
with the closed string charge. All resulted field theory are
$D=4,\,\netone$ gauge theory with fundamental chiral multiplet and
with (or without) adjoint chiral multiplet. At the probe limit the
known supergravity solution needs not to be modified. The effect
of fundamental flavors is manifested by the induced DBI action on
worldvolume of probing branes. If probing branes do not carry any
induced charges, their stability is ensured by the negative mass
modes in 5d spacetime. In particular, adding spacetime filling
D5$'$-branes in the MN background makes the background
supergravity dual to a large $N$ SQCD with non-chiral quarks.

We can introduce two distinct sets of probing D5-branes into the
original wrapped D5 background. The three sets of D5-branes are
orthogonal to each other. Then this configuration gives rise to a
$SU(M_1)\times S(M_2)$ global symmetry, as in QCD. It is a
non-chiral global symmetry. When we introduce mass deformation
mixing two distinct flavors, for $M_1=M_2$ the global symmetry in
the field theory gets broken to the diagonal one. It corresponds
to two orthogonal 5-branes merging into one smooth 5-branes
wrapping on a 2-cycle\cite{AKLM00,AK00}. A unsolved difficulty is
how to incorporate chiral symmetry via the above approach. The
usual treatment is to use Hanany-Witten setups\cite{HW96} in brane
setups\cite{BH97}. The setups, however, have to introduce NS5
branes and a background with singularity such as
orbifold\cite{PRU00} and consequently it destroies the original
wrapped D5-brane background.

We have pointed out that the worldvolume theories of probe branes
include effects of fundamental flavor. Authors of
refs.\cite{meson1,meson2} argued that, for a spacetime filling
brane probe, the worldvolume theory is just a meson theory. The
meson masses and mass gap has been studied in the AdS case. In
principle it can be extended to the MN background. The difficulty
is that the MN background are much more complicated than the AdS
background, and there is no explicit R-symmetry to classify the
meson spectrum.

\appendix

\section{Harmonic function on a 5-cycle inside a CY 3-fold}

A negative mass mode in the AdS$_{d+1}$ space does not lead to
instability as long as the mass is above the BF bound\cite{BF82a},
$(mR)^2\geq -d^2/4$ with curvature radius $R$. In scenario of
AdS$_5$/CFT$_4$ correspondence, the negative mass mode is from
Kaluza-Klein spectrum of the IIB supergravity on AdS$_5\times
S^5$\cite{GM85,KRN85}. For example, eigenvalues of the scalar
harmonic function on $S^5$ are $-l(l+4)$ and they are exactly
related to the spectrum in AdS$_5$ as $(mR)^2=l(l+4)\geq -4$. In
the MN background, the masses of particles propagated in 5d bulk
spacetime are also determined by the Kaluza-Klein spectrum of the
IIB supergravity on a compact 5-cycle inside a CY 3-fold. In other
words, we need to evaluate eigenvalues of harmonic functions on
the 5-cycle. Because this 5-cycle possesses normal bundle
structure, it is very difficult to evaluate all eigenvalues of
harmonic functions. In this paper we only focus on the existence
of positive eigenvalues (negative masses). Precisely, there is no
usual eigenvalue for harmonic functions on the 5-cycle since
metrics on a 5-cycle depends on the radius parameter $\rho$. Thus
we should call them as eigenvalue-function of $\rho$. It works as
a potential in 5d bulk spacetime. We will show, however, there is
a constant part in the eigenvalue-function of scalar harmonic
function, which generate masses of particles propagated in 5d bulk
spacetime.

Let us consider IIB equation of motion for the scalar field $H$ in
10d, \beq{a1}\na^2_{10}H=\frac{1}{\sqrt{-\hat{g}}}\pa_M
 (\hat{g}^{MN}\pa_N H)=\na^2_{bulk}H+\na^2_{5}H=0,
\eeq
with
\beq{a2}\na^2_{bulk}=\alpha'g_sN\eta^{\mu\nu}\pa_\mu\pa_nu+
  \frac{e^{-4\phi}}{\sqrt{g_5}}\frac{\pa}{\pa\rho}
  (e^{4\phi}\sqrt{g_5}\frac{\pa}{\pa\rho}),
\eeq is the Laplace operator defined by bulk metric in
eq.~(\ref{3.1}) and
\beq{a3}\na^2_5=\frac{1}{\sqrt{g_5}}\pa_i(g^{ij}\sqrt{g_5} \pa_j),
\eeq is defined by 5-cycle metric
\beq{a4}ds_5^2=e^{2h}(d\theta_1^2+\sin^2{\theta_1}d\phi_1^2)
     +\sum_{a=1}^{3}(w^a-A^a)^2,
\eeq where $g_5=e^{4h}\sin^2{\theta_1}\sin^2{\theta_2}$ is the
determinant of this metric.

At UV limit ($\rho\to\infty$) the metric~(\ref{a4}) reduces to
\beq{a5}ds_5^2&\simeq &\rho(d\theta_1^2+\sin^2{\theta_1}d\phi_1^2)
     \nonumber \\
     &+&\frac{1}{4}\left[d\theta_2^2+\sin^2{\theta_2}d\phi_2^2
     +(d\psi+\cos{\theta_1}d\phi_1+\cos{\theta_2}d\phi_2)^2\right].
\eeq It has the product form of $S^2\times S^3$. Since radius of
$S^2$ goes to infinity, harmonic functions on $S^2$ give
continuous spectrum. The main contribution is from $S^3$ with
constant radius. The eigenvalue of scalar harmonic function on
$S^3$ is $-l(l+2)$. It indeed contains a positive eigenvalue 1.

At IR limit ($\rho\to\ 0$) the metric~(\ref{a5}) becomes
\beq{a6}
  ds_5^2&\simeq &\rho^2(d\theta_1^2+\sin^2{\theta_1}d\phi_1^2)
     +\frac{1}{4}(d\theta_1+\cos{\psi}d\theta_2+\sin{\psi}
      \sin{\theta_2}d\phi_2)^2 \nonumber \\
     &+&\frac{1}{4}(\sin{\psi}d\theta_2-\cos{\psi}\sin{\theta_2}d\phi_2
       +\sin{\theta_1}d\phi_1)^2 \nonumber \\
     &+&\frac{1}{4}d\psi+\cos{\theta_1}d\phi_1+\cos{\theta_2}d\phi_2)^2.
\eeq At $\rho=0$ the $S^2$ shrinks to zero and it is $S^3$ with
constant radius. At the leading order of $\rho^2$ expansion, the
explicit form of the Laplace operator on 5-cycle is \beq{a7}
\rho^2\na_5^2&\simeq &
 \frac{1}{\sin{\theta_1}}\frac{\pa}{\pa\theta_1}
  (\sin{\theta_1}\frac{\pa}{\pa\theta_1})+(\cot^2{\theta_1}
  +\cot{\theta_1}\cot{\theta_2}\cos{\psi})\frac{\pa^2}{\pa\psi^2}
  -\cos{\psi}\frac{\pa}{\pa\theta_1}\frac{\pa}{\pa\theta_2}
  \nonumber \\
  &+&2\sin{\psi}\cot{\theta_2}\frac{\pa}{\pa\theta_1}
  \frac{\pa}{\pa\psi}+\frac{1}{\sin^2{\theta_1}}
  \frac{\pa^2}{\pa\phi_1^2}-\frac{2\sin{\psi}}{\sin{\theta_2}}
  \frac{\pa}{\pa\theta_1}\frac{\pa}{\pa\phi_2} \nonumber \\
  &-&\frac{2}{\sin{\theta_1}}(\cot{\theta_1}+\cos{\psi}\cot{\theta_2})
  \frac{\pa}{\pa\psi}\frac{\pa}{\pa\phi_1}
  +\frac{\cos{\psi}}{\sin{\theta_1}\sin{\theta_2}}
  \frac{\pa}{\pa\phi_1}\frac{\pa}{\pa\phi_1} \nonumber \\
  &+&(\theta_1\leftrightarrow\theta_2,\,\phi_1\leftrightarrow\phi_2)
   +O(\rho^2).
\eeq Here we focus on the maximal non-zero eigenvalue of the above
operator. Two quantum number $m_1$ and $m_2$ are generated by two
$U(1)$ symmetries associating to the shift of $\phi_1$ and
$\phi_2$. The maximal eigenvalue of the Laplace corresponds to
$m_1=m_2=0$. Then harmonic function
$H=\sin{\theta_1}\sin{\theta_2}\sin{\psi}$ gives maximal
eigenvalue $-2$. Hence $\na_5^2$ at the leading order we yield an
attractive potential \beq{a8} \na_5^2\sim
-\frac{\lambda}{\rho^2},\hspace{1in} \lambda=0,2,\cdots. \eeq It
is obtained by shrinking $S^2$ wrapped by D5-branes to zero.

The negative mass mode in 5d bulk spacetime comes from the
sub-leading order of $\na_5^2$ whose eigenvalues are constant. In
fact, explicit expression of $\na_5^2$ in sub-leading order is
precisely the same as the Laplace on $S^3$. It agrees with our
first glimpse that the 5-cycle is equivalent to $S^3$ at the IR
limit. The eigenvalue of scalar harmonic function on a 5-cycle,
therefore, indeed contains a positive constant eigenvalue 1 at the
sub-leading order.

Matching UV result with IR result, we conclude that
eigenvalue-function of scalar harmonic function has the form
\beq{a9} -l(l+2)+V(\rho), \eeq where \beq{a10}
V(\rho)&=&\,\frac{a_1}{\rho}+\frac{a_2}{\rho^2}+\cdots,
\hspace{1in} \rho>1, \nonumber \\
V(\rho)&=&\frac{b_1}{\rho^2}+b_2\rho^2+\cdots,\hspace{0.9in}\rho<1.
\eeq This result indicates that there are indeed negative mass
modes in the 5d bulk parameterized by the MN background. They
associate to the Kaluza-Klein spectrum on $S^3$ in a 5-cycle. When
we introduce spacetime filling D-branes probe in the MN
background, they cancels the NS-NS tadpole and consequently it
ensure the stability of spacetime filling branes.

The authors greatly thank Prof. J.-X. Lu and M. Li for useful
discussions. The work is partially supported by the NSF of china,
10231050.

\end{document}